\documentclass[nature,floatfix,superscriptaddress]{revtex4}
\pdfoutput=1 

\usepackage{graphicx}
\usepackage{epstopdf}
\usepackage{amsmath}
\usepackage{hyperref}
\usepackage{upgreek}
\hypersetup{
    colorlinks=true,
    linkcolor=blue,
    citecolor=blue
}

\newcommand{\NIST}{
National Institute of Standards and Technology, Boulder, CO, USA}
\newcommand{\CU}{
Department of Physics, University of Colorado, Boulder, CO, USA}
\newcommand{\equal}{
\thanks{These authors contributed equally.}}

\newcommand{\Al}{Al$^+$ }
\newcommand{\Yb}{Yb }

\begin{document}

\title{Optical coherence between atomic species at the second scale: improved clock comparisons via differential spectroscopy}

\author{May E. Kim}
\equal
\affiliation{\NIST}

\author{W. F. McGrew}
\equal
\affiliation{\NIST}
\affiliation{\CU}

\author{N. V. Nardelli}
\equal
\affiliation{\NIST}
\affiliation{\CU}

\author{E. R. Clements}
\equal
\affiliation{\NIST}
\affiliation{\CU}

\author{Y. S. Hassan}
\affiliation{\NIST}
\affiliation{\CU}

\author{X. Zhang}
\affiliation{\NIST}
\affiliation{\CU}

\author{J. Valencia}
\affiliation{\NIST}
\affiliation{\CU}

\author{H. Leopardi}
\affiliation{\NIST}
\affiliation{\CU}

\author{D. B. Hume}
\email{david.hume@nist.gov}
\affiliation{\NIST}

\author{T. M. Fortier}
\email{tara.fortier@nist.gov}
\affiliation{\NIST}

\author{A. D. Ludlow}
\email{andrew.ludlow@nist.gov}
\affiliation{\NIST}
\affiliation{\CU}

\author{D. R. Leibrandt}
\email{david.leibrandt@nist.gov}
\affiliation{\NIST}
\affiliation{\CU}

\maketitle

% The subheadings are for internal use; they will not be used for the version that is submitted to Nature Photonics
\section{Abstract}

Comparisons of high-accuracy optical atomic clocks \cite{Ludlow2015} are essential for precision tests of fundamental physics \cite{Safronova2018}, relativistic geodesy \cite{McGrew2018, Grotti2018, Delva2019}, and the anticipated redefinition of the SI second \cite{Riehle2018}. % 28 words
The scientific reach of these applications is restricted by the statistical precision of interspecies comparison measurements. % 17 words
The instability of individual clocks is limited by the finite coherence time of the optical local oscillator (OLO), which bounds the maximum atomic interrogation time. % 25 words
In this letter, we experimentally demonstrate differential spectroscopy \cite{Hume2016}, a comparison protocol that enables interrogating beyond the OLO coherence time. % 19 words
By phase-coherently linking a zero-dead-time (ZDT) \cite{Schioppo2017} \Yb optical lattice clock with an \Al single-ion clock via an optical frequency comb and performing synchronised Ramsey spectroscopy, we show an improvement in comparison instability relative to our previous result \cite{network2020frequency} of nearly an order of magnitude. % 47 words
To our knowledge, this result represents the most stable interspecies clock comparison to date. % 14 words
% 150 words total

% this would be a nice ending to the abstract, but I don't know what to chop to make it fit: and demonstrates that comparison protocols such as differential spectroscopy provide new possibilities for applications of atomic clocks.

\section{Main}

Atomic clocks based on optical transitions in single trapped ions or ensembles of neutral atoms in optical lattices have achieved remarkable fractional systematic uncertainties at the $10^{-18}$ level or below \cite{Brewer2019, McGrew2018, Bothwell2019, Huang2021, Huntemann2016}, establishing them as the most accurate measurement devices of any kind.  Beyond their application in timekeeping, this unrivalled accuracy makes optical atomic clocks ideal sensors for a variety of scientific and technological applications.  For example, the most stringent constraints on models of ultralight dark matter \cite{network2020frequency, Lange2021} are set by frequency ratio measurements of clocks based on different atomic species, with correspondingly different sensitivities to proposed physics beyond the Standard Model \cite{Derevianko2016}.  In addition to high accuracy, these constraints and other sensing applications are enhanced by high interspecies clock comparison bandwidth and stability.

The fractional frequency instability of a clock based on Ramsey spectroscopy of $N$ uncorrelated atoms is fundamentally limited by quantum projection noise (QPN) at
\begin{equation}
    \sigma_{\text{QPN}} (\tau) = \frac{\text{e}^{\Gamma_{\text{c}} T /2}}{2\pi\nu_{\text{c}}\sqrt{NT\tau}} \ ,
    \label{eq:sigma}
\end{equation}
where $\Gamma_{\text{c}}$ is the excited-state decay rate of the clock transition, $\nu_{\text{c}}$ is the unperturbed atomic transition frequency, $\tau$ is the averaging time, and $T$ is the Ramsey free-evolution (interrogation) time. The QPN limit is particularly restrictive for single-ion clocks ($N = 1$), as opposed to lattice clocks which routinely operate with $N \approx 10^4$ atoms. However, singly and multiply ionised \cite{Kozlov2018} atoms and molecules \cite{Chou2020} offer a wider variety of transitions suitable for clock operation, some of which are highly sensitive to proposed effects beyond Standard Model  including dark matter couplings \cite{Banerjee2020} and fundamental symmetry violations \cite{Lange2021}.  Instability due to QPN can be reduced by extending the interrogation time, but for Ramsey spectroscopy OLO noise restricts $T$ to less than the coherence time of the laser \cite{Leroux2017}. For large $T$, the evolved relative phase between the atoms and the laser can wander outside the range [$-\pi/2,\pi/2$], beyond which phase measurements become ambiguous, and it is no longer possible to lock the frequency of the OLO to the atomic transition.  %To date, the vast majority of interspecies optical clock comparisons have used independent OLOs, limiting the maximum interrogation time of each clock to the coherence time of its OLO.

This limit to the stability of independent clocks has motivated the development of several novel clock comparison techniques in which the two clocks use phase-synchronised OLOs to enable interrogation durations beyond the laser coherence time. As was first proposed and demonstrated by \citet{Bize2000} in fountain clocks, by synchronising the interrogations of the two clocks and using the same OLO, it is possible to suppress common-mode OLO noise, though the total interrogation time remains limited by OLO decoherence \cite{Takamoto2015, Schioppo2017}. Using a protocol called correlation spectroscopy \cite{Chwalla2007}, it is possible to interrogate beyond the OLO coherence limit for differential measurements, as was demonstrated in a recent lifetime-limited comparison of two \Al clocks \cite{Clements2020}. Correlation spectroscopy is effective only for clocks similar in frequency, but other protocols are usable for more general comparisons. Such protocols can be realised by using one or multiple systems to prestabilise the OLO \cite{Borregaard2013, Schioppo2017}, by making non-destructive measurements to realise a phase lock \cite{Kohlhaas2015, Bowden2020}, or by employing dynamical decoupling \cite{Dorscher2020}.  \citet{Hume2016} proposed one such protocol called differential spectroscopy, in which phase measurements from one clock are fed forward to a second clock and used to disambiguate the phase measurements of the second clock, which is operated beyond the laser coherence limit.

\begin{figure*}[htpb]
    \centering
    \includegraphics[width=0.75\columnwidth]{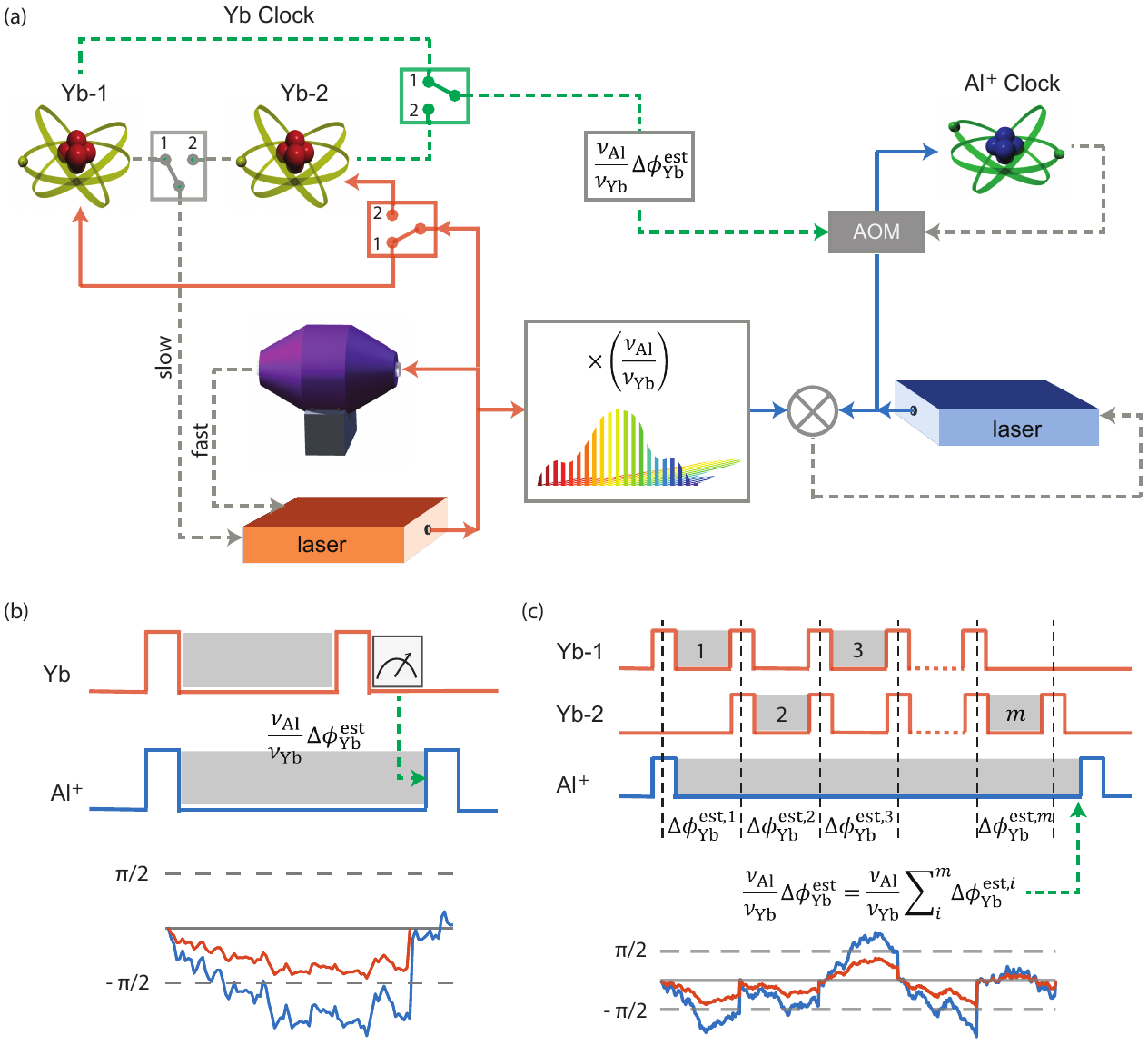}
    \caption{Experimental overview. (a) An ultrastable laser, prestabilised to an optical cavity, is used to interrogate a \Yb optical lattice clock. For ZDT measurements, two ensembles (Yb-1 and Yb-2) are interrogated in an interleaved fashion, and frequency feedback to the OLO and phase feedforward to the \Al clock are provided by the two ensembles, alternately. A frequency comb multiplies the \Yb OLO frequency up to near the \Al transition frequency, and the \Al OLO is phase locked to the comb. This OLO interrogates an \Al ion nearly synchronously with the \Yb interrogation, completing the interrogation soon after the \Yb phase feedforward is applied via an AOM to correct for common-mode laser noise. The grey dashed lines represent frequency feedback and the green dashed lines represent phase feedforward. The pulse sequence and a possible evolution of the integrated atom-laser phase during an interrogation are shown (b) for single \Yb ensemble measurements and (c) for ZDT measurements. The $\uppi/2$-pulses in Ramsey spectroscopy are depicted by the orange (blue) square pulses for the \Yb (\Al) clock. The grey shaded regions represent the Ramsey free-evolution periods. The two clocks start their interrogation at the same time, after which the relative phase between the OLO and the atoms evolves for the two clocks. After the \Yb clock makes a measurement of the atom-laser phase, the phase information is fed forward to the \Al clock so that it can complete the interrogation on the ion with the phase corrected by the \Yb clock. For ZDT measurements, the \Yb clock interrogates two \Yb optical lattice systems in an interleaved fashion, and after each interrogation, the OLO phase is adjusted for the \Al clock. The \Al clock completes its interrogation after $m$ cycles on the \Yb clock.}
    \label{fig:overview}%
\end{figure*}

In this work, we demonstrate a comparison between an ytterbium ($^{171}$Yb) optical lattice clock and a single-ion aluminium ($^{27}$Al$^+$) quantum-logic clock using differential spectroscopy. Figure \ref{fig:overview}a shows the experimental setup. The 267~nm \Al OLO is phase locked to the 578~nm \Yb OLO using an Er/Yb-doped glass femtosecond optical frequency comb. Maintaining phase coherence between the two systems via the frequency comb and active fibre noise cancellation at each experiment is critical for differential spectroscopy to be effective. The two clocks initiate Ramsey interrogation with synchronised $\pi/2$ pulses of 10~ms duration as shown in Fig.~\ref{fig:overview}b.  The evolved phase of the \Al clock during the Ramsey free-evolution time is given by $\Delta\phi_\text{Al} = \beta \Delta \phi_\text{Yb}$, where $\beta = \nu_\text{Yb}/\nu_\text{Al} \approx 2.16$ is the frequency ratio and $\Delta\phi_\text{Yb}$ is the evolved phase of the \Yb clock.  The measured \Yb phase is used to estimate $\Delta\phi_\text{Al}$, which is fed forward to correct the \Al OLO phase before its final Ramsey $\pi$/2 pulse. Because $\beta > 1$, the \Yb OLO coherence time is a factor of $\beta$ longer than that of the \Al coherence time. Assuming that \Yb has enough atoms to measure the phase with high signal-to-noise ratio and that the OLO instability is characterised by flicker frequency noise, the \Al interrogation time may be extended by up to a factor of $\beta$ longer than is possible without feedforward corrections from the \Yb clock.

We extend the original differential spectroscopy proposal by operating the \Yb clock in ZDT mode \cite{Schioppo2017}, in which two independent atomic ensembles are interrogated using Ramsey spectroscopy in an interleaved fashion.  Multiple interrogations with times $T_\text{Yb}$ are completed on the \Yb clock, with phase feedforward corrections applied to the \Al clock after each one, during a single \Al interrogation, as shown in Fig.~\ref{fig:overview}c. Further details about the experimental setup and protocol can be found in the Methods.  Using ZDT differential spectroscopy, the \Al clock interrogation time $T_\text{Al}$ may be extended yet further, potentially approaching the 20.6~s \Al excited state lifetime.

\begin{figure*}[htpb]
    \centering
    \includegraphics[width=0.9\columnwidth]{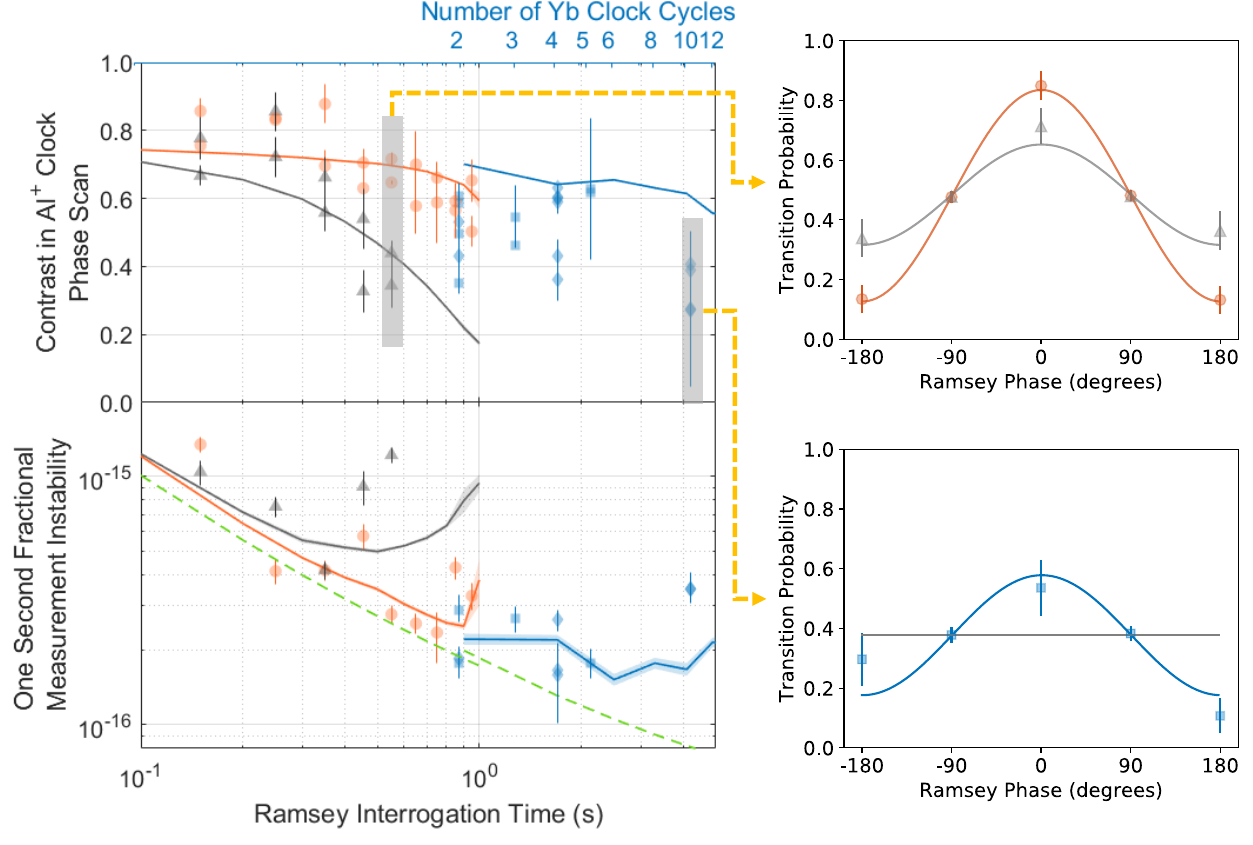}
    \caption{Differential spectroscopy clock comparison results for various $T_{\text{Al}}$ for the single \Yb ensemble case (orange circles) and varying the number of the \Yb clock interrogation cycles for the ZDT case (blue squares). Also shown are the results of spectroscopy without phase feedforward corrections (grey triangles). Numerical simulations are shown for single \Yb ensemble with corrections (orange line), single \Yb ensemble without corrections (grey line), and ZDT (blue line) cases for $m \in \{2,3,4,5,6,8,10,12\}$ with 1 sigma standard deviation error given by the shaded regions. The green dashed line represents the quantum projection noise limit of the ratio measurement instability. The boxes on the right show the Ramsey fringe (atomic transition probability as a function of the relative phase between the two Ramsey pulses) of the \Al spectroscopy with (orange curve for single \Yb ensemble and blue curve for the ZDT) and without (grey curve for both top and bottom) phase feedforward from the \Yb clock for two different values of $T_{\text{Al}}$. The maximum slope points are used for feedback to the AOM while the other points are interrogated once every 10 cycles to gather information on the contrast of the Ramsey fringe.}
    \label{fig:results1}%
\end{figure*}

\begin{figure*}[htpb]
    \centering
    \includegraphics[width=0.6\columnwidth]{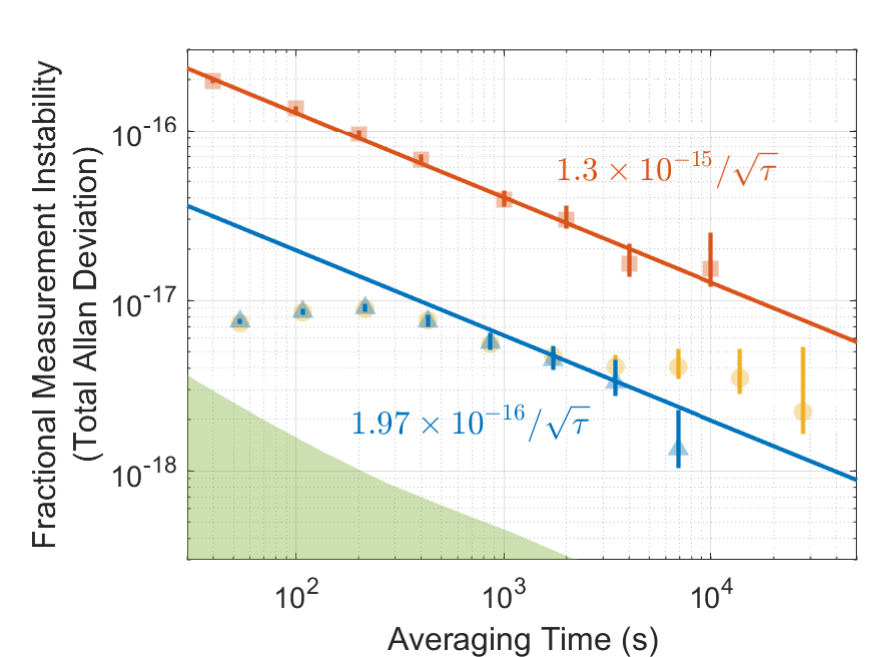}
    \caption{Stability of a 36 hour long ZDT differential spectroscopy clock comparison taken on 5-6 November 2020, with $m=4$ \Yb clock interrogations corresponding to a single \Al clock interrogation cycle for the full run (yellow circles) and an 8 hour selection of the total run (blue triangles). Our previous result compared \Yb and \Al (orange squares) using a conventional optical clock comparison technique. The solid lines are error-weighted, least-squares fits to the asymptotic region of the instability curves. The green shaded region represents a bound on the frequency transfer instability introduced by the frequency comb.}
    \label{fig:results2}%
\end{figure*}

\begin{figure*}[htpb]
    \centering
    \includegraphics[width=0.6\columnwidth]{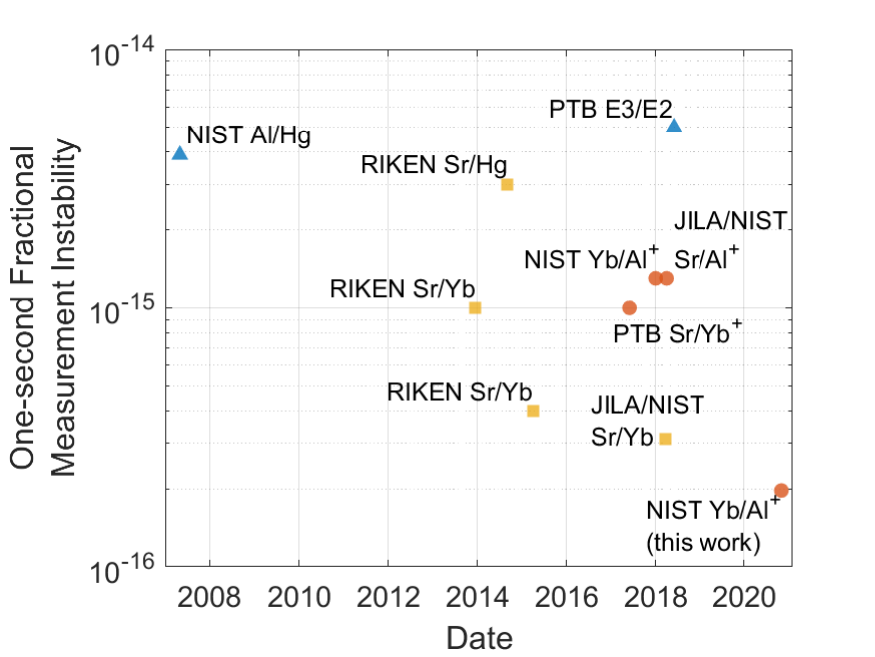}
    \caption{One-second instability of various interspecies optical clock comparisons plotted by their measurement date. Displayed are: comparisons between an optical lattice clock with another optical lattice clock (yellow squares), an ion clock with another ion clock (blue triangles), and an optical lattice clock with an ion clock (orange circles). Note that this is not an exhaustive list of all previous interspecies comparisons but does include the most competitive results. Each measurement is marked with the laboratories involved in the measurement, as well as the species of optical clocks participating \cite{Rosenband2008, Takamoto2015, Nemitz2016, network2020frequency,Dorscher2021,Huntemann2021}. }
    
    \label{fig:compare}%
\end{figure*}

The results of both single \Yb ensemble and ZDT experiments are shown in Fig.~\ref{fig:results1}. The single \Yb ensemble experiments are carried out for various values of $T_\text{Yb}$. We use $T_\text{Al} = T_\text{Yb} + 50$ ms to allow time for the \Yb clock to perform the second $\pi/2$-pulse, population measurement, and phase feedforward.  The ZDT experiments have a fixed $T_\text{Yb}=0.404$ s, which was chosen for safely interrogating the \Yb ensembles without a fringe hop, and the number of the \Yb clock interrogations is incremented. Also shown for comparison are the results of conventional Ramsey spectroscopy without phase feedforward. Each inset on the right of Fig.~\ref{fig:results1} shows the Ramsey phase scans of the \Al clock at a specific $T_\text{Al}$ value, one with feedforward corrections from the \Yb clock (orange and blue curves) and one without (grey curve), which are fit to determine the contrast. The power spectral density of the measured frequency ratio noise is fit to determine the long-averaging-time asymptotic Allan deviation (see Methods), which is shown in the bottom left panel of Fig.~\ref{fig:results1}. The improvement in the atom-laser phase coherence with corrections is shown by the increase in Ramsey contrast in both plots. As $T_\text{Al}$ increases, the interrogation time comes to exceed the coherence limit of the \Al OLO, and the Ramsey fringe without phase feedforward drops to zero contrast indicating that phase feedforward is required to run the \Al clock at these interrogation times.

The results of numerical simulations - which model comparisons between the two clocks with realistic noise generated to match the laboratory conditions - are also plotted for the single \Yb ensemble and ZDT cases and shown as solid lines in Fig. \ref{fig:results1}. With increasing $T_\text{Al}$, we expect the instability to improve, as can be seen by the QPN limit (green curve) plotted in the lower plot in Fig.~\ref{fig:results1}, reaching a minimal value of $5\times10^{-17}/\sqrt{\tau}$ at an interrogation time corresponding to the lifetime of the \Al clock's excited state. Instead, contrast in the Ramsey phase scan starts to diminish with the increase in $T_\text{Al}$, and the one-second instability reaches a minimum value approximately a factor of 3 above the fundamental limit of the \Al clock lifetime. Further simulation results indicate (see Methods) that technical noise sources in the system, such as the fibre links and local fluctuations in the magnetic field, compromise instability as $T_\text{Al}$ increases. The reasonable level of agreement between simulation results and empirical data suggests that, with further suppression of such sources of technical noise, it will be possible to obtain a comparison instability that approaches the lifetime limit of the \Al clock.

% changing uptime from 87.3 to 88.8
% actual duty cycle (with gain points, etc) changing from 83% to 67.9%

Figure \ref{fig:results2} (yellow circles) shows the Allan deviation of a 36-hour-long continuous ZDT differential spectroscopy comparison, with a duty cycle of 68$\%$ and a total uptime of 88.8$\%$, using four \Yb interrogations ($T_\text{Yb} = 0.404$~s) for each \Al interrogation ($T_\text{Al} = 1.72$~s). A drifting systematic frequency shift leads to an apparent $3 \times 10^{-18}$ flicker noise floor at averaging times greater than 30,000~s for differential spectroscopy. The source of this long-term noise has not yet been identified. For these measurements, we note that neither clock was operated in a configuration with a low systematic uncertainty (see Methods). However, an eight hour subset (blue triangles) shows that the data average as white frequency noise over this time scale, approaching the $1 \times 10^{-18}$ level. Furthermore, We anticipate that the apparent noise floor can be suppressed in future work by operating the clocks in low-uncertainty mode. The asymptotic instability of ZDT differential spectroscopy is $(1.97 \pm 0.11) \times 10^{-16}/\sqrt{\tau}$ (blue line), which is improved by nearly an order of magnitude over our previous result of $1.3\times 10^{-15}/\sqrt{\tau}$ using conventional techniques \cite{network2020frequency} (red line). % We note that an improvement of an order of magnitude in instability corresponds to a two orders of magnitude reduction in the total measurement time required to reach a target measurement precision.

For context, we plot the asymptotic instability of various interspecies comparisons in Fig. \ref{fig:compare} versus the date on which the comparison was performed. Due to the large number of atoms participating in each interrogation, optical lattice clocks generally have lower instability, and the most stable lattice-lattice comparisons (yellow squares) perform with lower instability than the most stable lattice-ion (orange circles) and ion-ion (blue triangles) comparisons. Nevertheless, due to our ability to push the duration of the Ramsey interrogation of the ion clock beyond the laser coherence time, our asymptotic instability is lower than even the best previous interspecies lattice-lattice clock comparisons. 

Differential spectroscopy improves the instability of clock comparison measurements by using a low-QPN clock to measure laser phase noise and perform feedforward corrections to a second clock's laser. This allows the second clock to operate with a longer interrogation time, beyond the coherence time of its laser, when the frequency ratio between the two clocks is greater than unity. Furthermore, by running the first clock in ZDT mode, the duration of interrogation of the second clock can be extended yet further, potentially approaching the excited-state lifetime limit. Though we demonstrate the technique for an interspecies comparison with a large frequency ratio, we note that ZDT differential spectroscopy can be utilised to improve other clock comparisons, even if the two clocks have the same transition frequency. This scheme can be understood as fusing the two clocks into a single, compound optical clock, capable of realising an instability that could be below that of either system taken independently. Lastly, the differential spectroscopy algorithm is also applicable for a second clock based on multiple entangled atoms or ions, which would allow clock comparison measurements at the Heisenberg instability limit.

We thank Roger C. Brown and Chin-wen Chou for their careful reading and feedback on this manuscript.  We also acknowledge contributions from Philip Rich, in carrying out a characterisation of the magnetic field noise of the Yb system. This work was supported by the National Institute of Standards and Technology, the Defense Advanced Research Projects Agency (Atomic-Photonic Integration Program), the National Science Foundation Q-SEnSE Quantum Leap Challenge Institute (Grant Number 2016244), and the Office of Naval Research (Grant Numbers N00014-18-1-2634 and N00014-20-1-2431).  M.E.K.~was supported by an appointment to the Intelligence Community Postdoctoral Research Fellowship Program at NIST administered by ORISE through an interagency agreement between the DOE and ODNI.  H.L.~was supported by the Department of Defense (DoD) through the National Defense Science \& Engineering Graduate (NDSEG) Fellowship Program.  The views, opinions, and/or findings expressed are those of the authors and should not be interpreted as representing the official views or policies of the Department of Defense or the U.S.~Government.

All authors contributed to the design of the experiment, collection of data, and revision of the manuscript.  During the measurements, Al$^+$ clock operation was conducted by E.R.C., D.B.H., M.E.K., D.R.L., and J.V.; Yb clock operation was conducted by W.F.M., Y.S.H., X.Z., and A.D.L.; comb metrology laboratory operation was conducted by T.M.F., H.L., and N.V.N.  Data analysis and preparation of the manuscript were performed by M.E.K., D.R.L., W.F.M., and N.V.N.

\section{Methods}

\subsection{Experimental setup}
% More detailed version of Fig. 1a

The two ytterbium optical lattice systems make use of one-dimensional optical lattices, operated near the magic wavelength, to confine the atoms in the Lamb-Dicke regime and perform spectroscopy of the ultranarrow clock transition, $^1\mathrm{S}_0\rightarrow {^3\mathrm{P}_0}$. An amplified external cavity diode laser (ECDL) at 1156 nm is frequency doubled to reach the 578 nm clock transition.  Both systems operate with approximately 10,000 atoms. The systems have been described in detail in \cite{McGrew2018}. A previous analysis of the instability for a zero-dead-time clock found that a one-second instability well below $1\times10^{-16}$ was accessible for these systems \cite{Schioppo2017}.

The \Al optical atomic clock used in this experiment has been described previously in \cite{Clements2020}. Spectroscopy is performed on the $^1\mathrm{S}_0\rightarrow {^3\mathrm{P}_0}$ clock transition of a single \Al ion. Sympathetic cooling and readout are achieved by co-trapping a single $^{40}$Ca$^+$ ion and using the motional coupling between the two ions \cite{Schmidt2005}. A fibre laser at 1068 nm is frequency doubled twice to deliver 267 nm light to the aluminium ion. The AOM in the path of this light is utilised for the feedback corrections from the aluminium ion and another for the feedforward operation in differential spectroscopy, which is discussed at length in Sec. \ref{sec:feedforward}. 

\begin{figure*}[htpb]
    \centering
    \includegraphics[width=0.8\columnwidth]{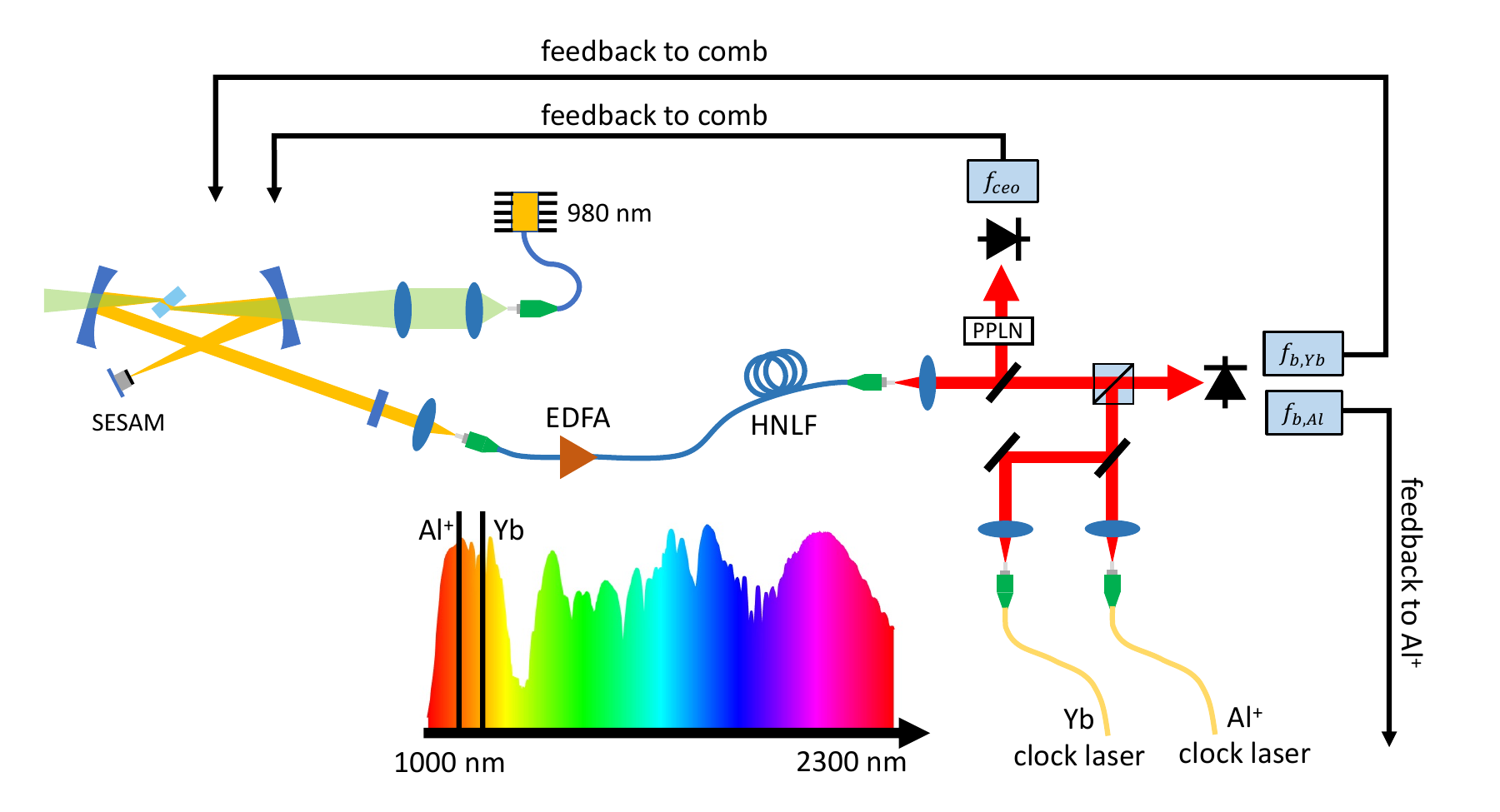}
    \caption{Optical paths surrounding the optical frequency comb portion of the experiment that enables coherent transfer of phase between the \Yb and \Al clocks.}
    \label{fig:comb}%
\end{figure*}

% How the clocks are connected
Phase-coherent transfer from the \Yb clock OLO at 1156 nm to the \Al clock OLO at 1068 nm is enabled by an optical frequency comb (OFC) shown in Fig.~\ref{fig:comb}. We use a self-referenced OFC based on an Er/Yb-doped glass gain medium that is pumped with 980 nm light and emits at 1550 nm, situated in a free-space cavity with 500 MHz free-spectral range. A semiconductor saturable absorption mirror (SESAM), which preferentially passes high-intensity light and absorbs low-intensity light, acts to mode-lock the laser and provides stable pulse formation. The mode-locked laser is laid out as a folded linear cavity with an asymmetry such that the beam is focused in the gain crystal as well as onto the SESAM end mirror. This ensures that the intra-cavity intensity is sufficient to minimize SESAM losses at the saturation and to maximize nonlinear self-phase modulation in the gain medium. The very low loss of the cavity (around 2 \%) yields a very narrow intrinsic free-running laser linewidth of less than 5 kHz. This is imperative for the low-noise phase transfer between clocks.

Pulses output from the laser cavity are amplified by an Er-doped fibre amplifier from about 70 mW to 300 mW and subsequently sent through a highly-nonlinear fibre (HNLF) for access to the carrier-envelope offset frequency, which requires spectral power on the two ends of an octave. We use a hybrid nonlinear fibre, comprising two stages: the first section (10 cm with dispersion 5.6 ps/(km $\cdot$ nm)) creates a dispersive wave around 1000 nm, and the second section (50 cm with dispersion 2 ps/(km $\cdot$ nm)) extends the spectrum to around 2100 nm. Careful dispersion management before the HNLF and tuning of the input optical pulse power allows us to put enough spectral power at 1068 nm and 1156 nm for detection of optical beats with the \Al and \Yb OLOs, respectively. A periodically-poled Lithium Niobate (PPLN) waveguide doubles light at 2070 nm and the resultant 1035 nm light is beat against the 1035 nm part of the spectrum for detection of the carrier-envelope offset frequency. The single-branch OFC design \cite{Leopardi2017} ensures that no out-of-loop fibre paths add instability to the comb.

The OFC is locked to the \Yb clock OLO by comparing the beat between the OLO and comb light to a 10 MHz H-maser reference and feeding back to the mode-locked laser cavity length via a fast piezo-electric transducer. The actuator is glued between the SESAM, which is very light, and a heavy lead-filled copper mount designed specifically to suppress resonances below several hundred kHz \cite{briles2010}. This affords comb noise suppression with a bandwidth of almost 200 kHz. A beatnote between a comb mode and light from the \Al clock OLO is fed back to the \Al clock lab to lock the OLO to the comb. To minimize instability due to out-of-loop optical free-space paths, the light from both OLOs was combined on a dichroic mirror, then interfered with the comb light on a single beam splitter and detected on a single photodiode. In this setup, there is only 20 cm of uncompensated optical free-space paths.

The \Al clock OLO is stabilised to a high-finesse optical cavity with high-bandwidth (around 300 kHz) feedback. When running differential spectroscopy, we use the phase of the beatnote between the comb mode and the \Al clock OLO as the error signal for low-bandwidth (around 100~Hz) feedback to lock the cavity-stabilised OLO to the comb.

\subsection{Differential spectroscopy feedforward corrections}
For differential spectroscopy, the 10-ms-duration of the first Ramsey $\uppi$/2-pulse are synchronised in the clocks at the several $\upmu$s level. The synchronisation is achieved by triggering the Ramsey pulse sequence on the \Al clock with a TTL signal sent from the \Yb clock experimental control system.

The second Ramsey $\uppi$/2-pulse is applied on the \Yb clock before the \Al clock to allow time for the \Yb clock to measure the phase wander of the shared OLO and feed it forward to the \Al clock. The total interrogation time of the atoms,  $T_{\text{Yb}}$, is set to be a fraction $\eta$ of the coherence time of the OLO at the \Yb transition frequency. To optimise $\sigma_{\text{QPN}}$ in Eq. \ref{eq:sigma}, the duration of the interrogation is maximised while ensuring that the relative phase between the OLO and the atoms, given by \cite{Hume2016}
\begin{equation}
    \Delta\phi_{\text{Yb}}=2\pi\int_0^{T_{\text{Yb}}}[\nu_{\text{Yb}}-f_{\text{Yb}}(t)]dt = 2\pi (\nu_{\text{Yb}}-\bar{f}_{\text{Yb}}) T_{\text{Yb}} \ ,
\end{equation}
does not exceed the range $[-\pi/2,\pi/2]$ (the so called invertible region) for Ramsey spectroscopy, where $\nu_{\text{Yb}}$ is the \Yb atomic transition frequency, $f_{\text{Yb}}$ is the time-dependent \Yb OLO frequency, and $\bar{f}_{\text{Yb}}$ is the mean frequency of the \Yb OLO during the clock interrogation.

After the second Ramsey $\uppi$/2-pulse on the \Yb clock, the ground state and excited state atom numbers are read out by destructive fluorescence detection using the strong ${^1}{\text{S}}_0 \rightarrow {^1}{\text{P}}_1$ transition at 399 nm. The frequency of the \Yb clock is locked to the centre of the resonance, but it is dithered between the half-maximum points on either side of the resonance centre, leading to an $\approx$ 50\% chance of excitation. Following this, the excitation ratio $r$ is determined, and the estimated optical phase of the OLO is calculated as
\begin{equation}
\Delta\phi_{\text{Yb}}^\mathrm{est}=\sin^{-1}(2(r-r_0)/A),
\end{equation}
where $r_0 \approx 0.5$ and $A \approx 1$ are the empirically determined half-maximum ratio and contrast of the Ramsey fringe, respectively. The \Yb OLO is frequency-locked to the \Yb atoms using the estimated deviation of the laser frequency  $\Delta\phi_\text{Yb}^\mathrm{est}/(2\pi T_{\text{Yb}})$ as the error signal. 

The \Al OLO, which is phase-locked to the \Yb OLO, interrogates the \Al ion for a duration $T_{\text{Al}} = T_{\text{Yb}}+ \tau_{\text{delay}}$ for a single \Yb ensemble experiment and $T_{\text{Al}} \approx mT_{\text{Yb}} + \tau_{\text{delay}}$ for a ZDT experiment - where $\tau_{\text{delay}}$ is the time required for the \Yb clock to make a measurement of its atom-laser phase and send the feedforward signal to the \Al clock, and $m$ is a positive integer. The final $\uppi$/2-pulse of the \Yb clock, atom detection, phase calculation, and phase correction are accomplished in several tens of milliseconds, but we choose to extend the free-evolution-time of the \Al clock by $\tau_{\text{delay}}=50$ ms to ensure that the phase correction has been applied before the second $\uppi$/2-pulse of the \Al clock. For ZDT mode, phase corrections are sent after each \Yb clock interrogation. The second $\uppi$/2-pulse of the \Al clock is supplied 50 ms after the second $\uppi$/2-pulse of the final ZDT cycle of the \Yb clock.

\begin{figure*}[htpb]
    \centering
    \includegraphics[width=0.75\columnwidth]{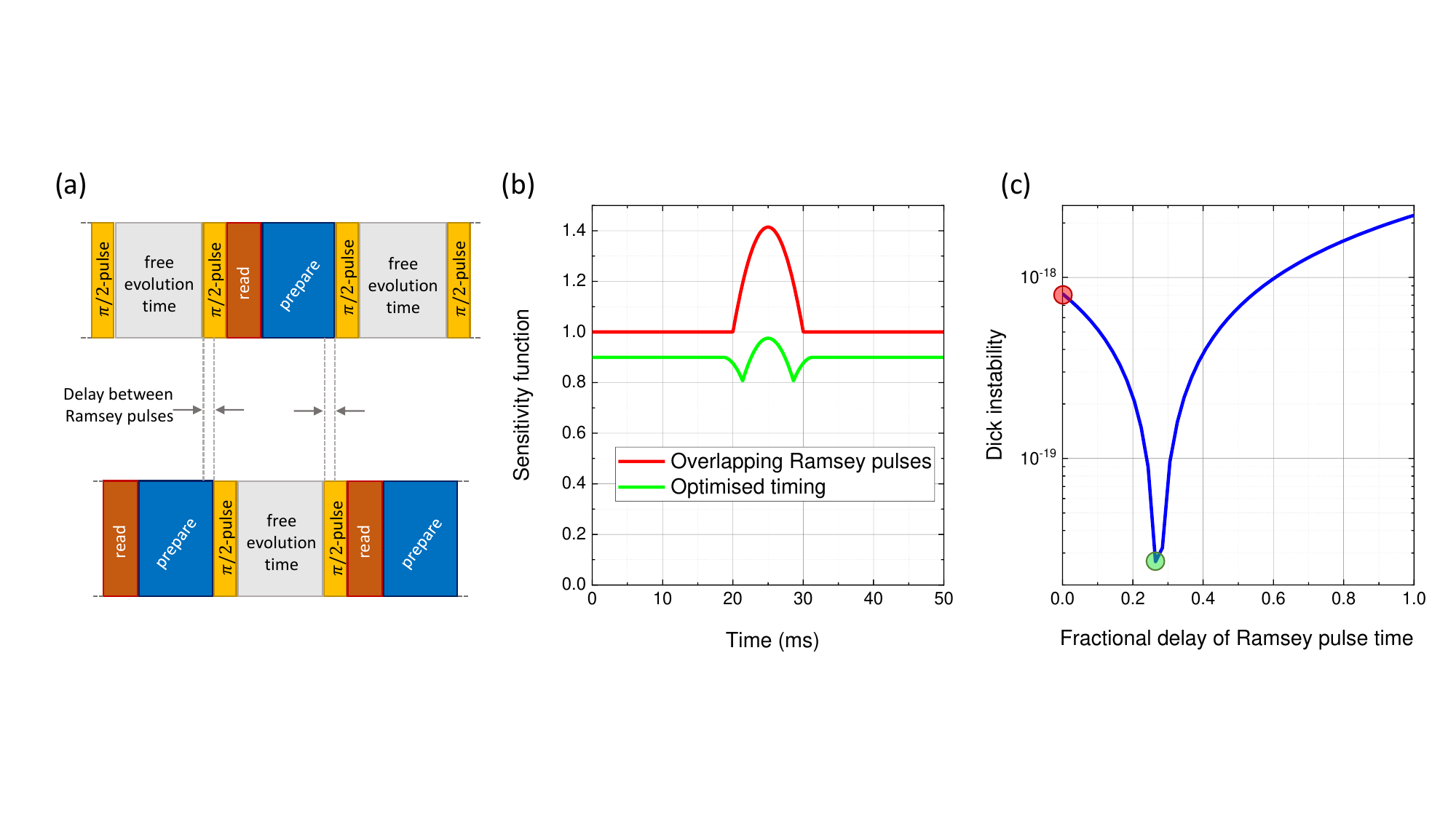}
    \caption{An optimised zero-dead-time clock. (a) The timing sequence for a ZDT clock with a variable delay between the $\uppi$/2-pulses. Note that the length of each step is not to scale. (b) Displayed is the sensitivity function for a zero-dead-time sequence in which the Ramsey pulses of the two atomic systems are precisely overlapped (red) or offset from each other by 27.73\% (green). The sensitivity functions have been vertically offset from each other for visual clarity. (c) The optimal delay is determined by calculating the one-second Dick instability for a variable delay between the two pulses. These calculations assume the OLO noise is flicker-frequency and at the level of $1.5\times10^{-16}$. The blue line demonstrates that the Dick instability exhibits a sharply defined minimum. We emphasise that the Dick instability reported here would be added in quadrature with other sources of instability, such as atomic detection noise and quantum-projection noise, which under current conditions would limit the total clock instability at a level at least an order of magnitude higher than the calculated Dick instability. The red and green dots represent the conditions with overlapping Ramsey pulses and optimised timing, respectively.}
    \label{fig:optimizedramsey}%
\end{figure*}

In Ref.~\cite{Schioppo2017}, we utilised a ZDT sequence in which the $\uppi$/2 pulses of the two systems overlapped with each other. Here, we use a slightly modified sequence: by adding a delay between the second $\uppi/2$-pulse of one of the systems and the first $\uppi/2$-pulse of the other, it is possible to further decrease the time-dependence of the ZDT clock's sensitivity function, and thus  decrease the Dick effect as well \cite{dick1990local}. Fig.~\ref{fig:optimizedramsey}(a) displays the sensitivity functions of a ZDT clock with simultaneous $\uppi/2$-pulses and a ZDT clock with a delay of 27.73\% of the $\uppi/2$-pulse time. Fig.~\ref{fig:optimizedramsey}(b) displays the calculated Dick effect as a function of delay time between the $\uppi/2$-pulses. In this work, we elect to operate our ZDT system with a delay of 2.773 ms, at which the calculated Dick effect is minimised.

The \Yb clock feeds forward the accumulated atom-laser phase after the termination of each of its interrogations weighted by the ratio of the two clock frequencies, $\beta$. The phase correction is sent via a serial connection to a 10 MHz direct digital synthesizer (DDS) in the \Yb lab which can be programmed with a 14-bit phase tuning word, implying a 0.4 mrad phase resolution for the corrections. The 10 MHz DDS serves as the external reference for a 160 MHz synthesizer in the \Al lab driving an AOM that frequency and phase shifts the \Al OLO just before it is delivered to the ion. Due to the difference of transition frequency between the clocks, the phase estimate must be multiplied by the frequency ratio $\beta$ such that $\Delta\phi_\text{Al}^\mathrm{est} = \beta \Delta\phi_{\text{Yb}}^\mathrm{est}$. Because the phase correction signal is multiplied by (160 MHz)/(10 MHz) = 16, the phase correction sent to the DDS must be further reduced by a factor of 16, $\Delta\phi^\mathrm{DDS} = \Delta\phi_{\text{Al}}^\mathrm{est}/16$.

Feeding forward the accumulated atom-laser phase after every \Yb interrogation is equivalent to feeding forward the estimated total atom-laser phase accumulated during the \Al clock interrogation time, $\Delta\phi_{\text{Yb,tot}}^{\text{est}}$. For the \Al clock, the relative phase between the OLO and atoms is given by
\begin{equation}
    \Delta\phi_{\text{Al}} = 2\pi(\nu_{\text{Al}}-\bar{f}_{\text{Al}})T_{\text{Al}} - \beta \Delta\phi_{\text{Yb,tot}}^{\text{est}}
\end{equation}
where the second term is the feedforward correction from the \Yb clock, $\nu_{\text{Al}}$ is the \Al atomic transition frequency, and $\bar{f}_{\text{Al}}$ is the mean frequency of the \Al OLO during the clock interrogation. For traditional operation of the \Al clock, the atom-laser phase is given simply by the first term, and $T_{\text{Al}}$ must be restricted such that $|2\pi(\nu_{\text{Al}}-\bar{f}_{\text{Al}})T_{\text{Al}}| < \pi/2$. However, for differential spectroscopy, the factor $\beta$ allows the interrogation time $T_{\text{Al}}$ to be increased by the same factor; and in the case of ZDT differential spectroscopy, the constraint on $T_{\text{Al}}$ is lifted such that excursions of the atom-laser phase far beyond the invertible region can be tolerated. This leads to a reduction in instability by $\sqrt{\beta m}$, as can be seen from Eq.~\ref{eq:sigma}.

After the second Ramsey $\uppi$/2-pulse on the \Al clock, the state of the \Al ion is read out using quantum-logic spectroscopy and the atom-laser phase estimate $\Delta\phi_{\text{Al}}^{\text{est}}$ is computed.  The \Al clock OLO is frequency-locked to the \Al ion by shifting the drive frequency of the same AOM used for feedforward phase corrections, $f_\mathrm{AOM}$, using the frequency estimate $\Delta\phi_{\text{Al}}^{\text{est}}/(2\pi T_2)$ as the error signal, with a feedback update rate of $1/T_{\text{Al}}$. The measured frequency ratio of the two clocks is thus
\begin{equation}
    \beta^{\text{est}} = \beta_0 + \frac{f_\mathrm{AOM}}{\nu_{\text{Yb}}} \ ,
\end{equation}
where $\beta_0$ is a known constant set by the mode numbers and frequencies of the comb beat notes and the frequencies of the other AOMs that shift the OLOs of both clocks between the comb and the atomic ensembles.

The raw frequency shift data from the two clocks are collected separately on each of the control computers. The Meinberg Network Time Protocol (NTP) programs loaded on the computers enable the aluminium clock control computer's onboard clock to synchronise to that of the ytterbium computer, such that the timestamps on the clock data collected on the two computers agree to within a few milliseconds. This enables us to look for noise that is correlated in both systems.

\label{sec:feedforward}

\subsection{Sources of noise in differential spectroscopy}
\subsubsection{Noise of the OLO}
Because the \Al clock laser is phase locked to the OLO of the \Yb clock, the noise characteristics of the \Yb clock OLO determine the durations of both clock interrogations. We measure its noise characteristics by measuring the performance of its optical cavity reference, and also by referencing the atoms. The former determines the spectrum of relatively high frequency noise, whereas the latter allows low frequency noise to be determined. Combining the results from these two spectra, we find that the \Yb clock OLO has a flicker noise limited fractional frequency noise of approximately $1.5\times 10^{-16}$ from 0.6~s to 1000~s. Below 0.6~s, the spectrum is approximately white frequency noise, and above 1000~s, the spectrum is approximately random walk/run. The performance of the OLO supports $T_{\text{Yb}} \approx 0.5$ s; but for ZDT mode differential spectroscopy, we use a conservative value of $T_{\text{Yb}} \approx 0.404$ s.

\subsubsection{Magnetic field noise}
Along with the phase noise fluctuations of the OLO, any environmental factors that can vary and induce uncertainty of the atomic transitions must be evaluated. One important source of differential instability is uncorrelated fluctuations of the magnetic field sampled by the clocks. Each clock accounts for the magnetic field fluctuations in its local environment by alternately probing the atoms on opposite Zeeman transitions: the \Yb atoms are interrogated on the $m_F = 1/2 \rightarrow m_{F'} = 1/2$ or $m_F = -1/2 \rightarrow m_{F'} = -1/2$ transitions and the \Al ion is interrogated on the $m_F = 3/2 \rightarrow m_{F'} = 1/2$ or $m_F = -3/2 \rightarrow m_{F'} = -1/2$ transitions. The latter is a departure from the traditional way of operating the \Al clock using the stretch states ($m_F = 5/2 \rightarrow m_{F'} = 5/2$ or $m_F = -5/2 \rightarrow m_{F'} = -5/2$) which minimises the sensitivity to magnetic field noise and extends the atomic dephasing time beyond the interrogation durations used in this work. The measured frequency difference between the opposite Zeeman transitions is used to determine the strength of the magnetic field for each system and is then fed back to the programmed difference between the interrogation frequencies used for opposite lines. %In the case of the \Yb clock running in ZDT mode, there are two \Yb systems operating in physically separated laboratory space such that the atoms in the two systems are subject to distinct magnetic fields. It takes four measurements (on two Zeeman lines on \Yb 1 and two on \Yb 2) to gather sufficient information on the magnetic field noise on the two systems. Thus, operating the \Al clock with cycles of \Yb clock that is not an integer multiple of 4 ($m \neq 4a$ where $a \in 1,2,3,...$) risk incorrectly attributing the source of the noise seen in the signals of the two Yb ensembles to the OLO instead of the magnetic field. This could introduce error in the estimate of the atom-laser phase that is fed forward to the \Al clock and lead to an increase in instability in differential spectroscopy. This risk is higher for operating the clock with two \Yb ensembles than a single \Yb ensemble, which does not suffer from a differential magnetic field noise. 

% The magnetic field noise in each system is measured using fluxgate magnetometers placed along the periphery of the science chamber where the atoms/ion are/is trapped, as well as by using atoms to measure the drift in the magnetic field. The noise components of the magnetic field is mostly flicker and random walk. 

In the \Yb clock, the atomic signal is used to servo the OLO laser frequency between the Zeeman sublevels. Due to the relatively long 1/e servo time constant of 44 s, we use a fluxgate magnetometer to measure the magnetic field fluctuations at shorter times. With the combination of these two data sets, we determine that the rms value of the noise in the \Yb clock is $4 \times 10^{-5}$ G.

In the \Al clock, the magnetic field is actively stabilised by feeding forward  fluxgate magnetometer measurements to the three orthogonal pairs of Helmholtz coils that surround the science chamber. The measurement details are discussed in \cite{Clements2020}. The rms value of the magnetic field noise in the \Al clock is $5 \times 10^{-6}$ G.

\subsubsection{Fibre link noise}
One source of noise that is important in differential spectroscopy is the frequency transfer uncertainty arising from imperfect phase-noise-cancellation (PNC) of the clock light distribution among the participating laboratories. The determination of the atom-laser phase in the \Yb clock is critical to the success of differential spectroscopy. However, if the frequency transfer system introduces dephasing by some $\Delta\phi_\text{PNC}$, then the estimated atom-laser phase $\Delta\phi_\text{Al}^{\text{est}} = \Delta\phi_\text{Al} - \Delta\phi_{\text{PNC}}$  will no longer be a good estimate of $\Delta\phi_\text{Al}$.

To mitigate this noise source, we use a path-length stabilisation technique \cite{Ma1994} for each fibre link between the \Yb clock to the \Al clock. At the receiving end of each fibre, a flat facet reflects a portion of the laser power back through a bidirectional acousto-optic modulator (AOM) to the sending end. The beatnote between the reference arm and the returned light is stabilised by feeding back to the drive rf of the AOM.

\begin{figure*}[htpb]
    \centering
    \includegraphics[trim = 1.5in 2.75in 1.5in 3in, clip, scale=0.6]{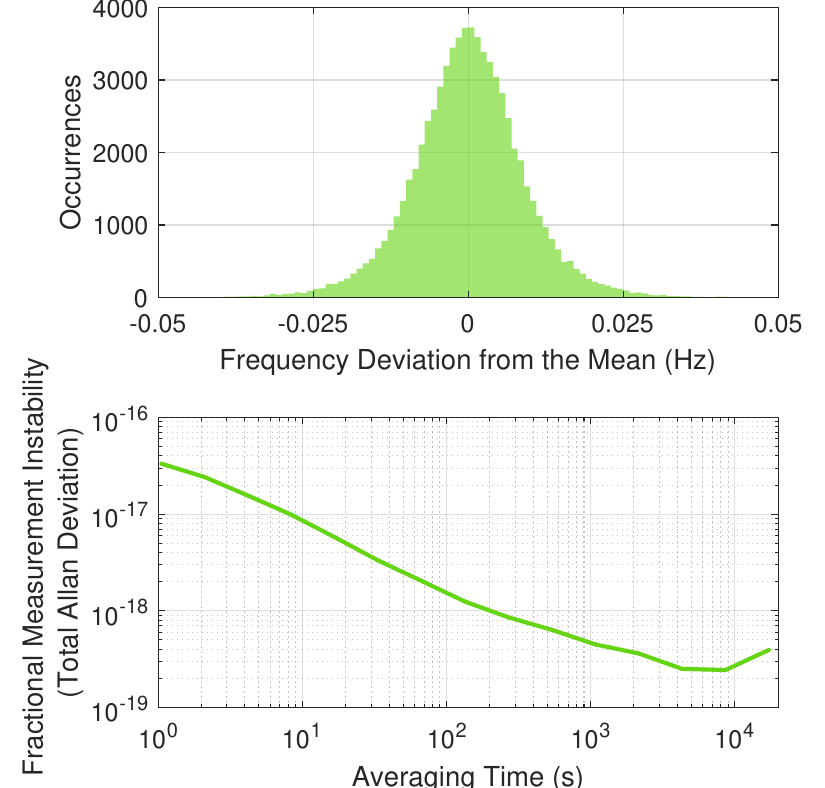}

    \caption{Instability added by phase-stabilised fibre network. The \Al clock OLO, which is locked to the \Yb clock OLO through the Er/Yb:glass frequency comb, and the \Yb clock OLO are compared with a Ti:Sapphire frequency comb. The top figure shows the histogram of the frequency noise, and the bottom shows the Allan deviation plot.}
    \label{fig:loopback}%
\end{figure*}

Additive uncertainty in the optical network is determined with a loopback measurement. Light from the \Yb clock and the \Al clock OLOs is sent over phase-stabilised links to the Er/Yb:glass frequency comb, which locks the \Al clock OLO to the \Yb clock OLO. Both OLOs also send light over stabilised fibre links to an independent Ti:Sapphire \cite{Fortier2006} frequency comb to measure the instability shown in Figure \ref{fig:loopback}. We measure a total network instability of $3.3 \times 10^{-17}$ at 1 second, which is approximately twice the instability experienced by the path from the \Yb clock to the \Al clock. The optical frequency combs add negligible instability around $3 \times 10^{-18}$ at 1 second, limited by uncompensated free-space optical paths \cite{Leopardi2017}.

\subsubsection{Other sources of noise}
The \Yb clock systems were operated in a similar mode to that described in \cite{McGrew2018}, with two main alterations. Firstly, a relatively high Rabi frequency of $2\uppi/(40$ ms) was used, facilitating differential spectroscopy, but impacting the level of systematic effects from probe Stark shift, line pulling, and AOM phase chirp. Second, some windows for the blackbody radiation shield of Yb-2 (described in \cite{Beloy2014}) were not in place; we estimate that this could potentially result in drifts in the blackbody radiation shift at the $1 \times 10^{-18}$ level. With these alterations, the systematic uncertainty reported in \cite{McGrew2018} was not valid for these runs, and further investigation would be required to ensure long-term stability at that level. Our ZDT algorithm allows us to measure the fractional frequency difference between the two atomic systems during a run, and we note that no drift was observed at the $2 \times 10^{-18}$ level between Yb-1 and Yb-2 during the run reported in Fig. 3. In any case, these sources of systematic uncertainty can be reduced. Fundamentally, differential spectroscopy scheme should be compatible with uncertainty budgets at the $10^{-18}$ level or below.

The systematic uncertainties of the \Al clock have not been fully evaluated, but it has been built to mitigate the leading cause of uncertainty in trapped-ion clocks, which is Doppler shift due to excess micromotion \cite{Brewer2019}, among other improvements over the previous generation \Al clock. One of the possible sources of uncertainty in this clock is the first-order Doppler shift due to slow drift of the ion position.  This can be caused by, for example, charging of dielectric surfaces near the ion by photoelectrons generated by scattered laser light. The first order Doppler shift was suppressed in the previous \Al clock by operating with two counterpropagating clock beams: one of which interrogated the ion while the other was detuned by 100 kHz. The beam on resonance and the detuned beam were alternated every cycle.  In this work, only one interrogation direction was used, and thus there may be an uncompensated first-order Doppler shift.  Another potential source of noise that has not yet been evaluated is a phase chirp of the AOM being used to supply the feedforward corrections to the \Al clock.

% Discussion on Figure 3

\subsection{Clock comparison instability limit}
ZDT differential spectroscopy has the potential to extend the interrogation time of the second clock (\Al in this work) to the fundamental limit set by the excited clock state lifetime of the atoms (20.6~s for the \Al clock), providing a corresponding improvement in instability. This is due to the fact that the first clock's interrogation time is always within the coherence time of the laser, and the atom-laser phase is continuously measured and fed forward to the second clock. As the second clock's interrogation time increases, the number of ZDT interrogation cycles of the first clock for each interrogation of the second clock also increases. However, the results of our experiment in which we scan the number of \Yb clock cycles do not show an improvement in instability after $T_\text{Al} \approx 2$ s (see Fig. \ref{fig:results1}).  Here, we present numerical simulations of the comparisons in which we vary three different sources of noise - the OLO noise, the magnetic field noise, and the fibre link noise - revealing that they all contribute to the comparison instability.

For the numerical simulations, OLO noise, magnetic field noise, and fibre link noise are numerically generated to match the experimentally measured noise spectra. White, flicker, and random walk noise models are used to generate the OLO noise with a fractional frequency noise floor of $1.5\times 10^{-16}$. The white noise is below 0.6 s and random walk and random run are beyond 1000 s. The \Yb and \Al clocks are subject to the same OLO noise with the \Al clock seeing noise amplitude increased by a factor of $\beta$. The simulations treat the two \Yb clocks separately such that they experience different magnetic field noise. The magnetic field noise is mostly flicker, and the rms value of the \Yb clock is $4 \times 10^{-5}$ G (with a linear Zeeman shift of 199.516 Hz/G), and the rms value of the \Al clock is $5 \times 10^{-6}$ G (with a linear Zeeman shift of about 280 Hz/G). Besides the magnetic field noise and the OLO noise, the \Al clock is subject to the fibre link noise, which imprints additional noise on the estimated atom-laser phase. The fibre link noise is mostly flicker in nature with both phase and frequency components and adds maximum instability of $3 \times 10^{-17}$ at 1 second. 

Both clocks probe the points of maximum slope on the central Ramsey fringe for highest sensitivity, on each side of the opposite Zeeman lines, such that a total of four interrogations is required for full knowledge of the magnetic field and OLO phase. In addition to these four interrogation frequencies, for every tenth interrogation of the \Al clock we interrogate a randomly choosen extrema (the maximum or one of the two minima on either side) of the central Ramsey fringe. This allows us to determine the Ramsey contrast in real time to evaluate how well the feedforward is working. 

In simulations, we generate atom-laser phase estimates made by the \Yb clock in the presence of noise after each interrogation. In the ZDT case, the \Yb clocks make $m$ measurements of the phase, and the total phase is fed forward to the \Al clock. The \Al clock, subject to its own set of noise sources, makes an estimate of the atom-laser phase, ideally within the invertible region with the assistance of the feedforward signal. The simulation results are compared to the experimental data, as can be seen in Fig. \ref{fig:results1}. 

\begin{figure*}[htpb]
    \centering
    \includegraphics[trim = 1.4in 2.6in 1.4in 2.4in, clip, scale=0.6]{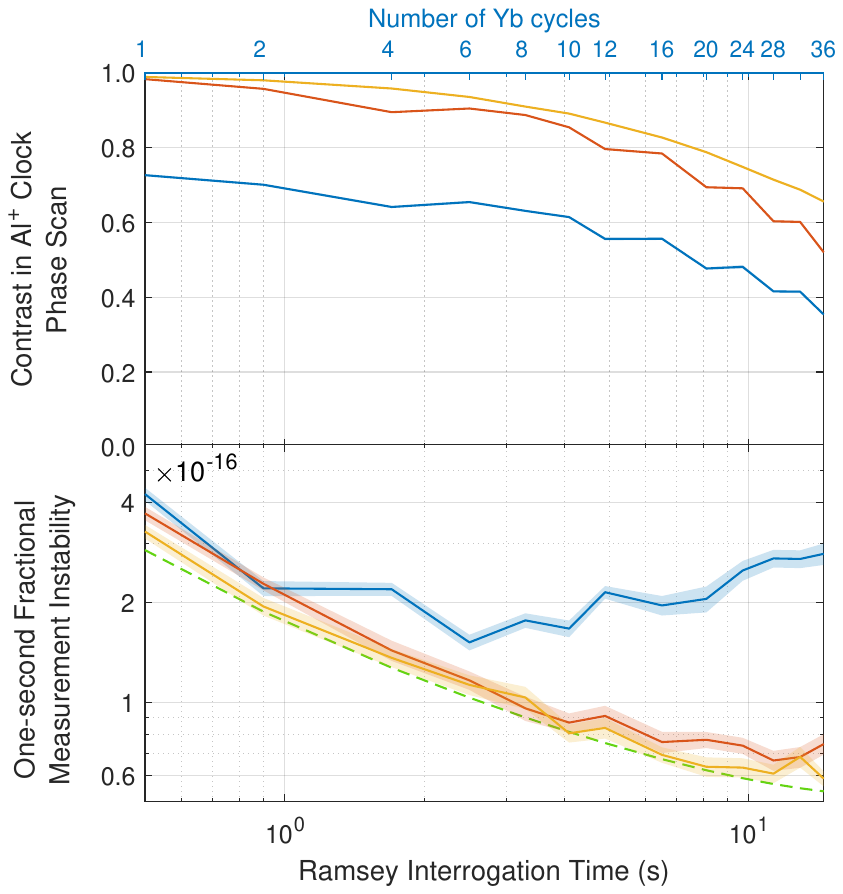}
    \caption{Results of the simulations that vary the noise sources in the operation of the ZDT differential spectroscopy operation. The top panel shows the contrast in the phase scan of the \Al clock, and the bottom shows the Allan deviation plot for: when only the OLO noise is present (yellow line), when OLO and the magnetic field noise sources are present (orange line), and when OLO, magnetic field, and the fibre link noises are present (blue line). The green dashed line in the bottom panel indicates the QPN limit. }
    \label{fig:simulated_noise}%
\end{figure*}

In order to explore the impact of the different noise sources on the clock comparison instability, we study three cases: when only the OLO noise is present, when the OLO noise is present along with the magnetic field noise, and when these two sources are present as well as the noise from the fibre link. Figure \ref{fig:simulated_noise} shows the contrast in the \Al clock's phase scan (top plot) and the fractional frequency instability (bottom plot) in the clock comparison for each of these three cases. The simulations incorporate the noise profiles of the OLO, fibre link, and the magnetic field in each system. The one second fractional measurement instability is extracted from each clock simulation run under various noise conditions. The statistical error associated with running 25 simulations for each Ramsey interrogation time is reflected in the shaded area. However, the error associated with extracting the one second instability is not reflected in Fig.~\ref{fig:simulated_noise}. This is responsible for the fact that the simulations curve falls below the QPN limit in Fig.~\ref{fig:results1}, which is not possible.

When the only source of noise present in the system is from the OLO, the instability continues to improve as $T_\text{Al}$ is increased out to several seconds, following the QPN limit. This indicates that without additional sources of noise, it would be possible to continue improving the clock comparison instability beyond 10 s. In our case, the limit on the Ramsey interrogation time - and hence the lowest achievable instability - is the excited state lifetime of the \Al clock transition, which is 20.6 s. If the lifetime of the excited state were infinite for the ion clock, the instability would be limited by the QPN limit of the \Yb clock. This would be reached when the number of clock cycles on \Yb clock is such that the phase noise accumulated reaches one radian, such that it is no longer contained in the invertible region of the phase space. A fringe hop would then be unavoidable, leading to higher instability. Due to the large number of atoms in the \Yb clock, many number of \Yb clock cycles will be necessary to lead to a fringe hop if no other noise source were an issue. 

When the magnetic field noise is present along with the OLO noise, the contrast in the \Al clock is somewhat reduced, and the instability increases slightly. The magnetic field is servoed in the \Al clock system and its RMS noise level is about $5\times 10^{-6}$~G, and the least magnetically sensitive transitions are used for the Ramsey interrogation. This allows the magnetic field noise to be suppressed in the clock comparison, even at Ramsey interrogation time above 10 s. A larger source of noise in the clock comparison arises from the fibre link between the various systems. when the fibre link noise is also present along with the OLO noise and the magnetic field noise, the clock comparison instability does not improve beyond $1.4\times 10^{-16}$ in fractional frequency, which is reached for $T_\text{Al}$ at about 2~s to 4~s. 

It is possible to improve the various sources of noise in our system to reach the quantum projection noise limit (Fig. \ref{fig:simulated_noise}), which continues to improve beyond $T_\text{Al} > 8$~s. Research is already underway to build a cryogenic optical cavity as a reference for the OLO to reduce its noise; and the \Al clock could be magnetically shielded to passively reduce the impact of the magnetic field noise from the environment. Improving the fibre link noise will allow the Ramsey interrogation time to increase, allowing us to achieve a lower instability. 

\subsection{Instability data analysis}

\begin{figure*}[htpb]
    \centering
    \includegraphics[width=1.0\columnwidth]{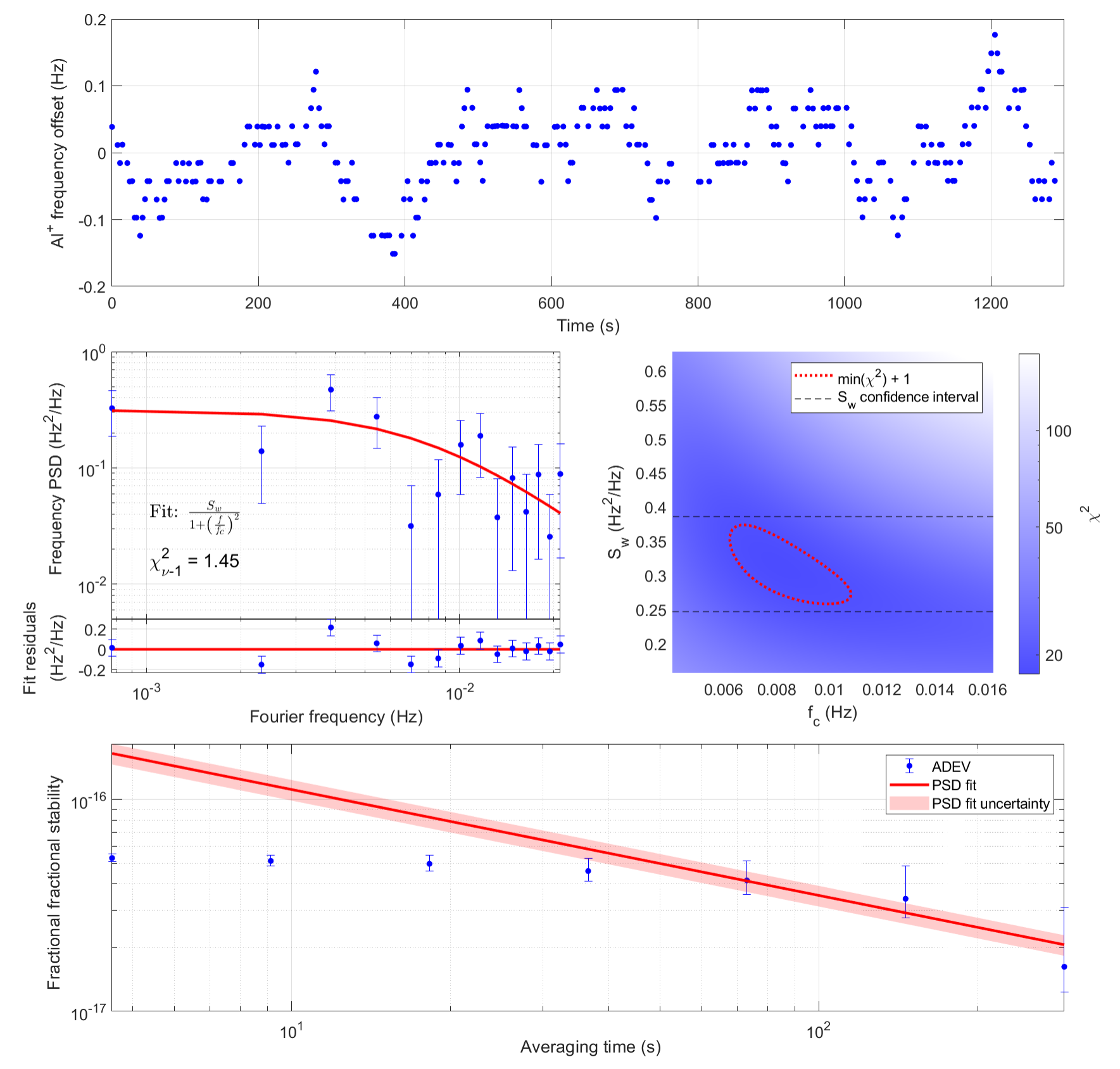}
    \caption{Example of the data analysis used to determine the long-averaging-time asymptotic Allan deviation.  This data set is Ramsey spectroscopy of the \Al clock without any feedforward corrections from \Yb, at a interrogation time of 454~ms.  The top panel shows the time series of the measured \Al transition frequency relative to the \Yb stabilised OLO.  The middle left panel shows the corresponding power spectral density of frequency noise, which is fit to the function $S_w/(1 + (f/f_c)^2)$.  The fit residuals are shown at the bottom of this panel.  The middle right panel shows $\chi^2$ as a function of the two fit parameters.  The red dotted line is a contour of constant $\chi^2 = \min(\chi^2) + 1$, and the black dashed lines are the 68~\% confidence interval bounds for the fit parameter $S_w$.  The bottom panel shows the Allan deviation of the frequency data (blue circles) and the long-averaging-time asymptotic Allan deviation corresponding to the $S_w$ fit parameter (red line and shaded confidence interval).}
    \label{fig:data_analysis}%
\end{figure*}

For each data set, the instability at long averaging times is determined by fitting the power spectral density of the measured frequency noise, an example of which is shown in the middle left panel of Fig.~\ref{fig:data_analysis}.  The fit function,
\begin{equation}
    S_f(f) = \frac{S_w}{1 + (f/f_c)^2} \ ,
\end{equation}
corresponds to white frequency noise of magnitude $S_w$ that has been low-pass filtered with a corner frequency of $f_c$.  This is physically motivated because the OLO is stabilised to the quantum-projection-noise limited \Yb clock and the \Al clock's frequency measurement is also quantum-projection-noise limited, resulting in a white frequency noise spectrum.  However, the frequency measurement bandwidth is limited by the frequency lock of the \Al clock to the OLO.  Because the \Al frequency lock uses a single frequency integrator, the expected measurement frequency noise PSD is low-passed white noise.  The white frequency noise fit parameter $S_w$ is converted into the long-averaging-time asymptotic Allan deviation $\sigma(\tau) = \sigma_{1s} / \sqrt{\tau}$ using $\sigma_{1s} = \sqrt{S_w/2} / \nu_\textrm{Al}$.

Notably, the two fit parameters $S_w$ and $f_c$ are correlated, as shown by the non-circular contour of $\chi^2$ in the middle right panel of Fig.~\ref{fig:data_analysis}.  The reduced $\chi^2$ of the fit in this example $\chi^2_{\nu-1} = 1.45$.  The upper and lower uncertainties of the fit parameter $S_w$ are determined as
\begin{equation}
    \sigma_{S_w,\textrm{upper}} = \sqrt{\chi^2_{\nu-1}} \times \left( \max_{\chi^2 = \min(\chi^2) + 1}(S_w)  - S_w \right)
\end{equation}
and
\begin{equation}
    \sigma_{S_w,\textrm{lower}} = \sqrt{\chi^2_{\nu-1}} \times \left( S_w - \min_{\chi^2 = \min(\chi^2) + 1}(S_w) \right) \ .
\end{equation}

For long data sets, this data analysis procedure gives results that are consistent with direct fits of the Allan deviation at long averaging times, which are typically reported.  However, for short data sets, it is often difficult to determine what range of averaging times should be included in such a fit, and the fit results depend strongly on this subjective choice.  Furthermore, the Allan deviation values at different averaging times are correlated with each other, making it difficult to determine the uncertainty of direct fits of the Allan deviation.  The procedure described here avoids these problems.

\medskip

\bibliography{main}

\end{document}